\documentclass[journal]{IEEEtran}
\usepackage{amsmath,amssymb,amsfonts}
\usepackage{algorithmic}
\usepackage{graphicx}
\usepackage{textcomp}
\usepackage{xcolor}
\usepackage{stfloats}
\usepackage{booktabs}

\newtheorem{assumption}{Assumption}[section]

\newtheorem{theorem}{Theorem}[section]

\newtheorem{lemma}{Lemma}
\newtheorem{remark}{Remark}

\begin{document}
%
\title{Data-Driven Controllability Analysis and Stabilization for Linear Descriptor Systems}

\author{Jiabao He, Xuan Zhang, Feng Xu*, Junbo Tan and Xueqian Wang
\thanks{J. He, F. Xu, J. Tan and X. Wang are with the Center for Intelligent Control and Telescience, Tsinghua Shenzhen International Graduate School, Tsinghua University, Shenzhen 518055, P.R.China (email: hjb18@tsinghua.org.cn; xu.feng@sz.tsinghua.edu.cn; tjblql@sz.tsinghua.edu.cn; wang.xq@sz.tsing-hua.edu.cn).}
\thanks{X. Zhang is with the Tsinghua-Berkeley Shenzhen Institute, Tsinghua University, Shenzhen 518055, P.R.China, (email: xuanzhang@sz.tsing-hua.edu.cn).}
\thanks{Corresponding author: Feng Xu.}
}


\maketitle

\begin{abstract}
For a parameter-unknown linear descriptor system, this paper proposes data-driven methods to testify the system's type and controllability and then to stabilize it. First, a data-based condition is developed to identify whether this unknown system is a descriptor system or is equivalent to a normal system. Furthermore, various controllability concepts are testified by replacing the descriptor system's matrices with data. Finally, a data-based decomposing method is proposed to transfer the nominal system into its slow-fast subsystems' form, so that a state feedback controller for the slow subsystem can be obtained from persistently exciting input and state sequences. Meanwhile, due to the equivalent stabilizability between the nominal system and its slow subsystem, a state feedback controller which stabilizes the nominal system is also obtained. A simulation example is provided to illustrate the effectiveness of those methods.
\end{abstract}


\IEEEpeerreviewmaketitle

\section{Introduction} \label{sec1}

From its birth to nowadays, model-based control (MBC) has experienced rapid development and updated many inspiring theories, for instance, from the classic control to modern control, and from the basic stability to optimal control and ${H_\infty }$ control \cite{willems2007in}. Based on mathematical models of systems, MBC theories are rigorous and interpretable, but when a control plant is a black box whose mathematical model is unavailable, MBC will lose its merits on rigor and interpretability to some extent. That is when we need to rely on data-driven control (DDC) which designs a controller directly from available on-line or off-line data. A detailed and early survey about MBC and DDC can be found in \cite{hou2013from}.

Recently, based on Willems et al.’s fundamental lemma in behavioral theory \cite{willems2005note}, researchers in \cite{de2019formulas} and \cite{van2020data} proposed a novel direct DDC method for linear time-invariant systems. This method parameterizes the system by persistently exciting input and corresponding state or output signals, so that there is no need to identify exact values of systems' parameters, and the stabilization and linear quadratic regulator (LQR) problems can be reformulated into data-dependent linear matrix inequalities (LMIs). Since the obtained controller parameters are interpretable under the framework of MBC, this method is more attractive than other DDC methods. Besides, researchers in \cite{van2020data} introduced the concept of data-informativity for systems' analysis and control. In order to improve the numerical feasibility of datasets and deal with missing data, the work in \cite{van2020Willems} extended the fundamental lemma to multiple datasets. Subsequent works have extended the DDC method in \cite{de2019formulas} to various scenarios, such as robust control \cite{berberich2020robust,bisoffi2021data,van2020noisy} and LQR \cite{dorfler2021certainty} in the presence of noisy data, data-based controllability tests \cite{mishra2020data}, linear time-varying systems \cite{nortmann2021direct}, switched linear systems \cite{rotulo2021online} and linear delay systems \cite{rueda2021data}. Moreover, model predictive control was considered in \cite{berberich2020data} and \cite{coulson2021distributionally}. More details and recent developments about DDC and the behavioral theory can be found in \cite{markovsky2021behavioral} and \cite{van2020fromdata}.

With no exception, all above methods assume that the investigated system is in the normal state-space form, i.e., the common $x_{k+1} = Ax_{k} + Bu_{k}$, where $x_{k}$ is the system state, $u_{k}$ is the control input, and $A$ and $B$ are unknown parameters. But there is a hidden problem. Since the system's model is unavailable, if we do not verify the system's type but directly exert inputs to the system, then the collected data for the subsequent analysis and control of an assumed normal system may be misused for other systems, in particular, under the circumstance of descriptor systems whose common form is $Ex_{k+1} = Ax_{k} + Bu_{k}$, where $E$ could be a singular matrix. Although researchers in \cite{wang2017data} firstly presented a data-based method to distinguish descriptor systems from normal systems, this method required the system state matrix $A$ to be non-singular, which was conservative and difficult to verify. Besides, comparing with fruitful research of MBC methods for descriptor systems \cite{duan2010analysis,dai1989singular}, there are few works and little attention on DDC methods for them. Since descriptor forms frequently appear in complex systems whose parameters are difficult to be obtained, such as electrical circuit systems, large-scale systems with interconnections and mechanical systems, it is necessary to research DDC methods for descriptor systems. Subspace model identification methods for descriptor systems were introduced in \cite{moonen1992subspace} and \cite{verhaegen1996subspace}, and reinforcement learning methods and iterative learning control for descriptor systems were presented in \cite{zhang2013data} and \cite{zhang2016data}, but all of them assumed that the unknown descriptor system had been well-posed into its slow-fast subsystems' form, rather than the nominal form. In practice, the first-hand data are generated from the nominal system rather than its slow-fast subsystems' form, i.e, it is natural and convenient for us to choose the nominal state and collect data, but it is difficult to access the state of the slow-fast subsystems. Thus, existing DDC methods for descriptor systems have strict assumptions, either on systems' forms or parameters, which makes them limited.

In a nutshell, it is necessary to identify systems' type, so that potential misuses of corresponding DDC methods can be avoided. Plus, due to the deficiency of DDC methods for descriptor system, this paper aims to propose a direct DDC method for a parameter-unknown linear descriptor system in the nominal form to testify the system's type and controllability and then to stabilize it. To be specific, this paper answers the following three questions for a descriptor system $E{x}_{k+1} = Ax_{k} + Bu_{k}$, where $E, A$ and $B$ are unavailable, i.e.,
\begin{enumerate}
  \item {Is it a normal system or a descriptor one?}
  \item {Is it controllable?}
  \item {How to design a state feedback controller to stabilize it?}
\end{enumerate}
For these purposes, three individual data experiments are devised, in which system input signals are well-designed and corresponding state signals are collected. Combining those data with related concepts in the MBC framework, we present data-based conditions to address the three questions above.

The main contributions are summarized as follows: (\uppercase\expandafter{\romannumeral1}) Unlike other related research, this paper removes unnecessary assumptions about descriptor systems' forms and parameters, except that descriptor systems should be regular and that their initial conditions should be consistent, which means our methods are general for systems analysis and controllers design. (\uppercase\expandafter{\romannumeral2}) A novel data-based decomposing method for descriptor systems is presented, which virtually transfers a nominal parameter-unknown descriptor system into its slow-fast subsystems' form. In this way, not only the system's properties, such as causality can be well revealed, but more importantly, due to equivalent stabilizability between the nominal descriptor system and its slow subsystem, a state feedback controller can be easily implemented for the slow subsystem, so that the nominal system is also stabilized.

The remainder of this paper is as follows: Section \ref{sec2} designs the first data experiment to testify the system's type, and it lays foundations for controllability analysis and controller design. Section \ref{sec3} proposes the second data experiment to analyze various controllability concepts. A persistently exciting data experiment is given in Section \ref{sec4} to design a state feedback controller. Section \ref{sec5} provides an example to illustrate the effectiveness of those methods. Finally, conclusions are presented in Section \ref{sec6}.

\emph{Notations:} $\mathbb{R}$ and $\mathbb{C}$ are the fields of real and complex numbers. For a matrix $X$, $\texttt{r}(X)$, $\texttt{det}(X)$, $\lambda(X)$, $\sigma(X)$, $X^T$, $X^{-1}$, $\|X\|_2$ and $X \succ 0$ mean the rank, the determinant, all eigenvalues, all singular values, the transpose, the inverse, the spectrum norm and the positive definiteness of $X$, respectively. The notation $\texttt{diag}(X,Y)$ means the diagonal matrix $\left[{\begin{array}{*{20}{c}}{X}&0\\0&{Y}\end{array}}\right]$, and $\lambda(X,Y)$ means all generalized eigenvalues of the matrix pencil $(X,Y)$. Besides, $I$ ($I_r$) and $0$ are identity and zero matrices of an appropriate dimension, and $\mathbf{1}$ is a constant vector in which all elements are 1.

\section{Data-Driven System Type Identification} \label{sec2}

Consider the following linear discrete-time descriptor system
\begin{equation}\label{E1}
    E{x}_{k+1} = Ax_{k} + Bu_{k}
\end{equation}
where $x_{k}\in \mathbb{R}^{n}$ is the system state, $u_{k}\in \mathbb{R}^{m}$ is the control input, $E, A \in \mathbb{R}^{n\times n}$ and $B\in \mathbb{R}^{n\times m}$ are unknown parameters. In order to testify the system's type, the rank of $E$ should be identified. Before introducing the first data experiment, we have the following necessary assumptions for the system \eqref{E1}.

\begin{assumption}[\cite{duan2010analysis}] \label{Ass1}
  The system \eqref{E1} is regular, i.e., there exists $s \in \mathbb{C}$, such that $\texttt{det} (s E - A)\neq 0$.
\end{assumption}

\begin{assumption}[\cite{duan2010analysis}] \label{Ass2}
  The initial state $x_{0}$ is consistent.
\end{assumption}

\begin{remark} \label{Rmk1}
  Unlike normal state-space systems which always have unique state solutions, descriptor systems will have no or not unique state solutions if Assumptions \ref{Ass1} and \ref{Ass2} are not satisfied. Thus, these two assumptions are necessary for analysis and control of descriptor systems, no matter by MBC or DDC methods. Besides, Assumption \ref{Ass1} implies that except for all generalized eigenvalues of the matrix pencil $(E,A)$, there are infinite numbers of $s \in \mathbb{C}$ such that $\texttt{det} (s E - A)\neq 0$ holds.
\end{remark}

Now, we introduce the first data experiment to distinguish between a normal system and a descriptor one.

\emph{\textbf{Experiment 1}}: Exerting $n$ groups of sub-experiments on the nominal system \eqref{E1}. For each sub-experiment, there are $l (l\geq 1)$ steps of input sequences which satisfy
\begin{equation} \label{E3}
\sum\limits_{k = 0}^{l - 1} {{u_k}(i)}  = 0, i = 1,2,...,n.
\end{equation}
Then for the $i\texttt{-th}$ sub-experiment, we have
\begin{equation} \label{E4}
\left\{ {\begin{array}{*{20}{l}}
{E{x_1}(i) = A{x_0}(i) + B{u_0}(i)}\\
{E{x_2}(i) = A{x_1}(i) + B{u_1}(i)}\\
\qquad \qquad \vdots \\
{E{x_l}(i) = A{x_{l - 1}}(i) + B{u_{l - 1}}(i)}.
\end{array}} \right.
\end{equation}
Based on \eqref{E3}, after summation on both sides of \eqref{E4}, we have
\begin{equation} \label{E5}
\sum\limits_{k = 1}^l {E{x_k}(i)}  = \sum\limits_{k = 0}^{l - 1} {A{x_k}(i)}.
\end{equation}
Since the system \eqref{E1} is regular, there must exist a scalar $s_0 \in \mathbb{C}$, such that $\texttt{det} (s_0 E - A)\neq 0$, then the equation \eqref{E5} can be rewritten as
\begin{equation} \label{E6}
E( {\left( {s_0  - 1} \right)\sum\limits_{k = 0}^{l - 1} {{x_k}(i)}  + {x_0}(i) - {x_l}(i)}) = \left( {s_0 E - A} \right)\sum\limits_{k = 0}^{l - 1} {{x_k}(i)}.
\end{equation}
Denoting vectors $n_i = {\left( {s_0  - 1} \right)\sum\limits_{k = 0}^{l - 1} {{x_k}(i)}  + {x_0}(i) - {x_l}(i)}$ and $m_i = \sum\limits_{k = 0}^{l - 1} {{x_k}(i)}$, then for all $n$ groups of sub-experiments, we have
\begin{equation} \label{E7}
EN = \left( {s_0 E - A} \right)M
\end{equation}
where $N = [ {{n_1}}\ {{n_2}}\ \cdots \ {{n_n}}]$ and $M = [
{{m_1}}\ {{m_2}}\ \cdots\ {{m_n}}]$. Since the matrix $N$ can always be observed and assigned to be non-singular, plus $\texttt{det} (s_0 E - A)\neq 0$, then $\texttt{r}(E) = \texttt{r}(M)$, so naturally we have the following theorem.

\begin{theorem} \label{Th1}
The descriptor system \eqref{E1} becomes a normal system if $\texttt{r}(M) = n$, and it is a descriptor system if $\texttt{r}(M) < n$.
\end{theorem}

\begin{remark} \label{Rmk2}
If $\texttt{r}(E) = \texttt{r}(M) = n$, then the system \eqref{E1} is equivalent to a normal system ${x}_{k+1} = E^{-1}Ax_{k} + E^{-1}Bu_{k}$, whose controllability analysis and stabilization have been well studied in \cite{de2019formulas,mishra2020data} and \cite{wang2011data}. Thus, we will mainly deal with the singular case, i.e., $\texttt{r}(E) < n$ in the rest of this paper.
\end{remark}

\begin{remark} \label{Rmk3}
The system \eqref{E1} is an ideal deterministic system, but noises will inevitably appear in the system. Here we consider two kinds of noises.

(\uppercase\expandafter{\romannumeral1}) Measurement noises: When the system states $x_{k}$ are measured by sensors, there will be noises generated by sensors, i.e., the system states will be $\hat{x}_{k} = x_k + d_{k}^{m}$, where $d_{k}^{m} \in \mathbb{R}^{n}$ are the measurement noises. In practical industrial processes, it is commonly assumed that the measurement noises have known statistical characteristics \cite{wang2011data}. Thus, the unbiased estimates of the real values of $x_{k}$ can be obtained, for instance, by repeating experiments. After replacing $x_{k}$ by its unbiased estimates, our method is still effective.

(\uppercase\expandafter{\romannumeral2}) System noises: If there are noises with zero mean in the system, then the system model will be $E{x}_{k+1} = Ax_{k} + Bu_{k} + d_{k}^{s}$, where $d_{k}^{s} \in \mathbb{R}^{n}$ are the system noises. For a large enough $l$ in each sub-experiment, we have $\sum\limits_{k = 0}^{l-1} {d_{k}^{s}} = \varepsilon \alpha$, where $\varepsilon$ is a scalar approaching 0 and $\alpha \in \mathbb{R}^{n}$ is a constant vector. Similarly, after lining up outcomes of the $n$ groups, we have
\begin{equation} \label{E7a}
EN = ({s _0}E - A)\hat M
\end{equation}
where $\hat M = M + \varepsilon {({s _0}E - A)^{ - 1}}\alpha \mathbf{1}^T$. Due to the small disturbance $\varepsilon {({s _0}E - A)^{ - 1}}\alpha \mathbf{1}^T$, the rank of $M$ will be changed, i.e., $\texttt{r}(E) \neq \texttt{r}(\hat M)$, so Theorem \ref{Th1} may lead to wrong conclusions. From a statistical perspective, a large $l$ means a small $\varepsilon $, so we have
\begin{equation} \label{E7b}
{\| {\hat M - M} \|_2} = | \varepsilon |{\| {{{({s _0}E - A)}^{ - 1}}\alpha \mathbf{1}^T}\|_2} \le \delta
\end{equation}
where $\delta$ is a scalar close to 0. According to the singular value decomposition (SVD) theory \cite{zhang2017matrix}, small disturbances have minor changes to matrices' singular values, i.e., $| {{\sigma _i(\hat M)} - {\sigma _i(M)}}| \le {\| {\hat M - M} \|_2} \le \delta$, where ${\sigma _i(\hat M)}$ and ${\sigma _i(M)}$ are the $i\texttt{-th}$ singular value of $\hat M$ and $M$. Supposing $\texttt{r}(E) = \texttt{r}(M) = r$, for $ i = r+1, r+2,...,n$, since ${\sigma _i(M)} = 0$, we have $|{{\sigma _i(\hat M)}}| \le \delta$. Due to the dynamics of the unstable system (1), when $l$ becomes larger, the values of elements in $\hat M$ become larger, which means that the distinction between the none-zero true singular values ${\sigma _1(\hat M)}, ... , {\sigma _r(\hat M)}$ and the perturbation ${\sigma _{r+1}(\hat M)}, ... , {\sigma _n(\hat M)}$ will become evident, so it is feasible for us to differentiate them. Thus, under the circumstance of system noises, we should compute the singular values of data matrix $\hat M$ rather than its rank. If there are singular values smaller than $\delta$, and their number is $r'$, then we have $\texttt{r}(E) = \texttt{r}(M) = n - r'$.
\end{remark}

\begin{remark} \label{Rmk4}
Besides the equation \eqref{E6}, we can have another equation based on \eqref{E5} as
\begin{equation} \label{E8}
\begin{split}
&( {s_0 E - A})\sum\limits_{k = 1}^l {{x_k}(i)}  \\
= &A( {\sum\limits_{k = 1}^l {( {s_0  - 1} ){x_k}(i)}  + s_0 ( {{x_0}(i) - {x_l}(i)} )} ).
\end{split}
\end{equation}
Denoting vectors $w_i =  {\sum\limits_{k = 1}^l {\left( {s_0  - 1} \right){x_k}(i)}  + s_0 \left( {{x_0}(i) - {x_l}(i)} \right)}$ and $v_i = \sum\limits_{k = 1}^l {{x_k}(i)}$, then for all $n$ groups of sub-experiments, we have
\begin{equation} \label{E9}
\left( {s_0 E - A} \right)V = AW
\end{equation}
where $V = [ {{v_1}}\ {{v_2}}\ \cdots \ {{v_n}}]$ and $W = [
{{w_1}}\ {{w_2}}\ \cdots\ {{w_n}}]$. Similarly, $W$ can always be non-singular. Although the equation \eqref{E9} has no help to the system's type identification, it will play an important role in controllability analysis, which we will show in the next section.
\end{remark}

\section{Data-Driven Controllability Analysis} \label{sec3}

For descriptor systems, their controllability concepts and conditions are more complex than normal systems. Usually, we need to consider the complete controllability (C-controllability), the reachable controllability (R-controllability) and the causal controllability (Y-controllability) \cite{dai1989singular,duan2010analysis}. The following experiment is designed to help us testify the system's controllability. The main idea is to replace the system's parameters with data, so that corresponding model-based controllability criteria can be expressed in a data-based way.

\emph{\textbf{Experiment 2}}: Exerting $m$ groups of sub-experiments on the  nominal system \eqref{E1}. In the $i\texttt{-th}$ sub-experiment, the input sequence is constant and chosen as
\begin{equation} \label{E10}
u_k^{'}(i) = {\left[{\begin{array}{*{20}{c}}{0,...,0,}&{\mathop{1}\limits_{i\texttt{-th}},}&{0,...,0}\end{array}} \right]^T}, i = 1,2,...,m.
\end{equation}
Then after $l (l\geq 1)$ steps of observations, we have $Ex_{l + 1}^{'}(i) = Ax_l^{'}(i) + Bu_l^{'}(i)$. After lining up outcomes of the $m$ groups, it comes to
\begin{equation} \label{E11}
E{R_1} = A{R_0} + B
\end{equation}
where ${R_1} = \left[ {\begin{array}{*{20}{c}}
{x_{l + 1}^{'}(1)}&{x_{l + 1}^{'}(2)}& \cdots &{x_{l + 1}^{'}(m)}
\end{array}} \right]$ and ${R_0} = \left[ {\begin{array}{*{20}{c}}
{x_l^{'}(1)}&{x_l^{'}(2)}& \cdots &{x_l^{'}(m)}
\end{array}} \right]$. Furthermore, choosing the same $s_0$ as in Experiment 1 such that $\texttt{det} (s_0 E - A)\neq 0$, then multiplying $(s_0 E - A)^{-1}$ to both sides of \eqref{E11}, we have
\begin{equation} \label{E12}
(s_0 E - A)^{-1}E{R_1} = (s_0 E - A)^{-1}A{R_0} + (s_0 E - A)^{-1}B.
\end{equation}
Finally, combining \eqref{E12} with \eqref{E7} and \eqref{E9}, we obtain
\begin{subequations}\label{E13}
\begin{align}
{\left( {s_0 E - A} \right)^{ - 1}}E &= M{N^{ - 1}} \tag{14a},\\
{\left( {s_0 E - A} \right)^{ - 1}}A &= V{W^{ - 1}} \tag{14b},\\
{\left( {s_0 E - A} \right)^{ - 1}}B &= M{N^{ - 1}}{R_1} - V{W^{ - 1}}{R_0}. \tag{14c}
\end{align}
\end{subequations}
For brevity, those data terms are denoted as $D_E := M{N^{ - 1}}$, $D_A :=V{W^{ - 1}}$, and $D_B :=M{N^{ - 1}}{R_1} - V{W^{ - 1}}{R_0}$. Note that the system's unknown parameters are all on the left side of \eqref{E13}, and all data are on the right side, so the rank of matrices $E, A$ and $B$ are well revealed by data matrices $D_E, D_A$ and $D_B$. By far, we have prepared well to analyze the system's controllability. In the rest of this section, model-based controllability conditions will be introduced first, then corresponding data-based conditions are provided.

\subsection{C-Controllability} \label{sec3.1}

\begin{lemma} [\cite{yang2004generalized}] \label{Lem1}
The descriptor system \eqref{E1} is completely controllable if and only if
\begin{equation} \label{E14}
\texttt{r}(\left[ {\begin{array}{*{20}{c}}
{sE - A}&B
\end{array}} \right]) = \texttt{r}(\left[ {\begin{array}{*{20}{c}}
{E}&B
\end{array}} \right]) = n, \forall s\in\mathbb{C}.
\end{equation}
\end{lemma}
\begin{lemma} \label{Lem2}
The descriptor system \eqref{E1} is completely controllable if and only if
\begin{equation} \label{E15}
\texttt{r}( {\left[ {\begin{array}{*{20}{c}}
{sI - E{{( {s_0 E - A})}^{ - 1}}}&B
\end{array}} \right]}) = n, \forall s\in\mathbb{C}.
\end{equation}
\end{lemma}

\emph{Proof:} (\uppercase\expandafter{\romannumeral1}) From Lemma \ref{Lem2} to Lemma \ref{Lem1}: If $s\neq 0$, then
\begin{equation}
\begin{split}
    n &= \texttt{r}( {\left[ {\begin{array}{*{20}{c}}
    {sI - E{{( {s_0 E - A})}^{ - 1}}}&B
    \end{array}} \right]} )\\
     &= \texttt{r}( {\left[ {\begin{array}{*{20}{c}}
    {sI - E{{( {s_0 E - A})}^{ - 1}}}&B
    \end{array}} \right]\left[ {\begin{array}{*{20}{c}}
    {{s^{ - 1}}( {s_0 E - A} )}&0\\
    0&I
    \end{array}} \right]})\\
     &= \texttt{r}( {\left[ {\begin{array}{*{20}{c}}
    {( {s_0  - {s^{ - 1}}})E - A}&B
    \end{array}} \right]} ) = \texttt{r}( {\left[ {\begin{array}{*{20}{c}}
    {s'E - A}&B
    \end{array}} \right]} )
\end{split}
\nonumber
\end{equation}
where $s' = ( {s_0  - {s^{ - 1}}})$; else if $s = 0$, then
\begin{equation}
\begin{split}
    n &= \texttt{r}( {\left[ {\begin{array}{*{20}{c}}
   {sI - E{{( {s_0 E - A})}^{ - 1}}}&B
   \end{array}} \right]} )\\
    &= \texttt{r}( {\left[ {\begin{array}{*{20}{c}}
   {sI - E{{( {s_0 E - A})}^{ - 1}}}&B
   \end{array}} \right]\left[ {\begin{array}{*{20}{c}}
   { - ( {s_0 E - A})}&0\\
   0&I
   \end{array}} \right]} )\\
    &= \texttt{r}( {\left[ {\begin{array}{*{20}{c}}
   E&B
   \end{array}} \right]} ).
\end{split}
\nonumber
\end{equation}
(\uppercase\expandafter{\romannumeral2}) A similar proof can be easily given to prove that Lemma \ref{Lem1} implies Lemma \ref{Lem2}, thus it is omitted.
\hfill $\square$

Lemma \ref{Lem2} shows that the C-Controllability of the descriptor system \eqref{E1} is equivalent to the controllability of the following normal system
\begin{equation}\label{E16}
 {\bar{x}}_{k+1} = A_E\bar{x}_{k} + B\bar{u}_{k}
\end{equation}
where $A_E = E{{( {s_0 E - A})}^{ - 1}}$.
Inspired by the controllability test of the normal system \eqref{E16}, we have the following data-based condition to analyze the C-Controllability of the descriptor system \eqref{E1}.

\begin{theorem} \label{Th2}
The descriptor system \eqref{E1} is completely controllable if
\begin{equation}\label{E17}
\texttt{r} ( {\left[ {\begin{array}{*{20}{c}}
{{D_B}}&{{D_E}{D_B}}& \cdots &{{{( {{D_E}})}^{n - 1}}{D_B}}
\end{array}} \right]} ) = n.
\end{equation}
\end{theorem}

\emph{Proof:} It is well known that the normal system \eqref{E16} is controllable if and only if
\begin{equation}\label{E18}
\texttt{r} ( {\left[ {\begin{array}{*{20}{c}}
B& A_EB& \cdots &{{{( A_E )}^{n - 1}}B}
\end{array}} \right]}) = n,
\end{equation}
which is equivalent to
\begin{equation} \label{E19}
\texttt{r} ( {{{( {s_0 E - A} )}^{ - 1}}\begin{bmatrix}B& A_EB&\cdots &{{{( A_E)}^{n - 1}}B} \\                                       \end{bmatrix}})= n.
\end{equation}
Based on \eqref{E13}, \eqref{E19} can be rewritten into the data-based form \eqref{E17}. Then due to the equivalent controllability between the normal system \eqref{E16} and the descriptor system \eqref{E1} in Lemmas \ref{Lem1} and \ref{Lem2}, the proof is completed. \hfill $\square$

\subsection{Y-Controllability} \label{sec3.2}

\begin{lemma} [\cite{yang2004generalized}] \label{Lem3}
The descriptor system \eqref{E1} is causal if and only if
\begin{equation} \label{E20}
\texttt{r}(\left[ {\begin{array}{*{20}{c}}
E&0\\
A&E
\end{array}} \right]) = n + \texttt{r}(E).
\end{equation}
\end{lemma}

\begin{theorem} \label{Th3}
The descriptor system \eqref{E1} is causal if
\begin{equation}\label{E21}
\texttt{r}(\left[ {\begin{array}{*{20}{c}}
{D_E}&0\\
{D_A}&{D_E}
\end{array}} \right]) = n + \texttt{r}(M).
\end{equation}
\end{theorem}

\emph{Proof:} According to \eqref{E13}, we have $\texttt{r}(\left[ {\begin{array}{*{20}{c}}
E&0\\
A&E
\end{array}} \right]) = \texttt{r}(\texttt{diag}(({s_0 E - A})^{ - 1},({s_0 E - A})^{ - 1})\left[ {\begin{array}{*{20}{c}}
E&0\\
A&E
\end{array}} \right])
= \texttt{r}(\left[ {\begin{array}{*{20}{c}}
{D_E}&0\\
{D_A}&{D_E}
\end{array}} \right])$.
Since $\texttt{r}(E) = \texttt{r}(M)$, we conclude that this theorem holds based on Lemma \ref{Lem3}. \hfill $\square$

\begin{lemma} [\cite{yang2004generalized}] \label{Lem4}
The descriptor system \eqref{E1} is causally controllable if and only if
\begin{equation} \label{E22}
\texttt{r}(\left[ {\begin{array}{*{20}{l}}
E&0&0\\
A&E&B
\end{array}} \right]) = n + \texttt{r}(E).
\end{equation}
\end{lemma}

\begin{theorem} \label{Th4}
The descriptor system \eqref{E1} is causally controllable if
\begin{equation}\label{E23}
\texttt{r}(\left[ {\begin{array}{*{20}{l}}
{D_E}&0&0\\
{D_A}&{D_E}&{D_B}
\end{array}} \right]) = n + \texttt{r}(M).
\end{equation}
\end{theorem}

\emph{Proof:} Based on the proof of Theorem \ref{Th3}, it is straightforward to obtain the results in Theorem \ref{Th4}. \hfill $\square$

\subsection{R-Controllability} \label{sec3.3}

\begin{lemma} [\cite{dai1989singular}] \label{Lem4r}
The descriptor system \eqref{E1} is R-controllable if and only if
\begin{equation} \label{E23r1}
\texttt{r}({W_M}) = {n^2}
\end{equation}
where ${W_M} \in \mathbb{R}^{{{n^2} \times (m + n)n}}$, and
\[
{W_M} = \begin{bmatrix}\begin{smallmatrix}
{ - A}&0&0& \cdots &0&0&B&0&0& \cdots &0&0\\
E&{ - A}&0& \cdots &0&0&0&B&0& \cdots &0&0\\
0&E&{ - A}& \cdots &0&0&0&0&B& \cdots &0&0\\
 \vdots & \vdots & \vdots & \ddots & \vdots & \vdots & \vdots & \vdots & \vdots & \ddots & \vdots & \vdots \\
0&0&0& \cdots &{ - A}&0&0&0&0& \cdots &B&0\\
0&0&0& \cdots &E&{ - A}&0&0&0& \cdots &0&B
\end{smallmatrix}\end{bmatrix}.
\]
\end{lemma}

\begin{theorem} \label{Th4r}
The descriptor system \eqref{E1} is R-controllable if
\begin{equation} \label{E23r2}
\texttt{r}({W_D}) = {n^2}
\end{equation}
where ${W_D} \in \mathbb{R}^{{{n^2} \times (m + n)n}}$, and
\[
{W_D} = \begin{bmatrix}\begin{smallmatrix}
{ - {D_A}}&0&0& \cdots &0&0&{{D_B}}&0&0& \cdots &0&0\\
{{D_E}}&{ - {D_A}}&0& \cdots &0&0&0&{{D_B}}&0& \cdots &0&0\\
0&{{D_E}}&{ - {D_A}}& \cdots &0&0&0&0&{{D_B}}& \cdots &0&0\\
 \vdots & \vdots & \vdots & \ddots & \vdots & \vdots & \vdots & \vdots & \vdots & \ddots & \vdots & \vdots \\
0&0&0& \cdots &{ - {D_A}}&0&0&0&0& \cdots &{{D_B}}&0\\
0&0&0& \cdots &{{D_E}}&{ - {D_A}}&0&0&0& \cdots &0&{{D_B}}
\end{smallmatrix}\end{bmatrix}.
\]
\end{theorem}
\emph{Proof:} After multiplying the non-singular matrix $\texttt{diag}({\left( {s_0 E - A} \right)^{ - 1}},...,{\left( {s_0 E - A} \right)^{ - 1}}) \in \mathbb{R}^{{n^2} \times {n^2}}$ to the left side of $W_M$ in \eqref{E23r1}, we obtain $W_D$ in \eqref{E23r2}. \hfill $\square$

\section{Data-Driven Stabilization} \label{sec4}

Data-based system analysis in the previous sections is indeed important, but as we emphasized before, DDC is more task-oriented, so a more important problem is how to design a controller based on data to command the system to perform as we expect. Among various control requirements, the stability is definitely the most important one. However, there is currently no direct DDC methods to stabilize the parameter-unknown system \eqref{E1}. In this section, we firstly present a novel data-based decomposing method which transfers the system \eqref{E1} into its slow-fast subsystems. Then a state feedback controller is designed from persistently exciting input and state sequences to stabilize the slow subsystem. In this way, the stability of the  nominal descriptor system is also guaranteed.

\subsection{Preliminaries} \label{sec4.1}

If the descriptor system \eqref{E1} is regular, then there exist two nonsingular matrices $Q\in \mathbb{R}^{n\times n}$ and $P\in \mathbb{R}^{n\times n}$ which transfer \eqref{E1} into its equivalent slow-fast subsystems' form:
\begin{subequations}\label{E2}
  \begin{equation} \label{E2a}
    x_{k + 1}^s = {A_s}x_k^s + {B_s}{u_k}, x_k^s\in \mathbb{R}^{n_1},
  \end{equation}
  \begin{equation} \label{E2b}
    N_fx_{k + 1}^f = x_k^f + {B_f}{u_k}, x_k^f\in \mathbb{R}^{n_2}
  \end{equation}
\end{subequations}
where ${n_1} + {n_2} = n$, $QEP = \texttt{diag}(I_{n_1},N_f)$, $N_f$ is nilpotent, $QAP = \texttt{diag}({A_s},I_{n_2})$, $QB = \left[{\begin{array}{*{20}{c}}{{B_s}}\\{{B_f}}\end{array}} \right]$, and $x = P\left[{\begin{array}{*{20}{c}}{x_k^s}\\{x_k^f}\end{array}} \right]$. The system \eqref{E2a} is known as a slow subsystem which is normal, and the system \eqref{E2b} is known as a fast subsystem. It should be pointed out that values of $Q$ and $P$ rely on the matrix pencil $(E,A)$. Although $Q$ and $P$ are not unique, there are implicit connections between each pair of them.
\begin{lemma} [\cite{dai1989singular}] \label{Lem5}
Supposing ${Q_1}E{P_1} = \texttt{diag}( {{I_{{n_1}}},N_{f_1}} )$, ${Q_1}A{P_1} = \texttt{diag}( {A_{s_1},{I_{{n_2}}}})$, and ${Q_2}E{P_2} = \texttt{diag}( {{I_{{n_1}}},N_{f_2}} )$, ${Q_2}A{P_2} = \texttt{diag}( {A_{s_2},{I_{{n_2}}}})$, where $N_{f_1}$ and $N_{f_2}$ are nilpotent, then there exist two nonsingular matrices $T_1\in \mathbb{R}^{n_1\times n_1}$ and $T_2\in \mathbb{R}^{n_2\times n_2}$, such that
\begin{equation}\label{E24}
{Q_1} = {\texttt{diag}({T_1},{T_2})}{Q_2},
{P_1} = {P_2}{\texttt{diag}({T_1},{T_2})}.
\end{equation}
\end{lemma}

Due to its decoupling form, the slow-fast subsystems are widely used for analysis and control of descriptor systems. In particular, the stabilizability of the nominal system \eqref{E1} and the slow subsystem \eqref{E2a} is equivalent.
\begin{lemma} [\cite{duan2010analysis}] \label{Lem6}
The descriptor system \eqref{E1} is stabilizable if and only if its slow subsystem \eqref{E2a} is stabilizable, i.e.,
\begin{equation} \label{E26}
\lambda (E,A + BK) = \lambda ({A_s} + {B_s}{K_s}) \subset {\Omega _{0,1}}
\end{equation}
where ${\Omega _{0,1}} = \{s|s\in\mathbb{C}, \left| s \right| < 1\}$ and
\begin{equation} \label{E26a}
K \in \left\{ {K|K = \begin{bmatrix}{{K_s}}&0\end{bmatrix}{P^{ - 1}},\lambda ({A_s} + {B_s}{K_s}) \subset {\Omega _{0,1}}} \right\}.
\end{equation}
\end{lemma}

For a sole normal system \eqref{E2a}, researchers in \cite{de2019formulas} presented the following direct DDC method to design a state feedback controller.

\begin{lemma} [\cite{de2019formulas}] \label{Lem7}
A controllable normal system \eqref{E2a} is persistently exciting of order $n+1$ if
\begin{equation} \label{E27}
\texttt{r}(\left[ {\begin{array}{*{20}{c}}
{{U_ {-} }}\\
{X_ {-} ^s}
\end{array}} \right]) = {n_1} + m
\end{equation}
where $T \ge (m + 1){n_1} + m$, $ {U_ - } = \begin{bmatrix}{u_0} & {u_1} & \cdots & {u_{T - 1}} \\\end{bmatrix}$ and $ {X_ - ^s} = \begin{bmatrix}{x_0^s} & {x_1^s} & \cdots & {x_{T - 1}^s} \\\end{bmatrix}.$
\end{lemma}

\begin{lemma} [\cite{de2019formulas}] \label{Lem8}
If the condition \eqref{E27} holds, then the normal system \eqref{E2a} is stabilized by a state feedback controller $u_k = K_{s}x_k^s$ if there exists a matrix ${\Phi _s} \in \mathbb{R}^{T\times n_1}$, such that
\begin{equation} \label{E28}
\left[ {\begin{array}{*{20}{c}}
{X_ - ^s{\Phi _s}}&{X_ + ^s{\Phi _s}}\\
{{{(X_ + ^s{\Phi _s})}^T}}&{X_ - ^s{\Phi _s}}
\end{array}} \right] \succ 0
\end{equation}
where $ {X_ + ^s} = \begin{bmatrix}{x_1^s} & {x_2^s} & \cdots & {x_{T}^s} \\\end{bmatrix}$, $X_ {-} ^s{\Phi _s} = {(X_ {-} ^s{\Phi _s})^T} \succ 0$, and the feedback gain is given by ${K_s} = {U_ {-} }{\Phi _s}{(X_ {-} ^s{\Phi _s})^{ - 1}}$.
\end{lemma}

\subsection{Controller Design} \label{sec4.2}

Now we introduce our third data experiment which helps to design a state feedback controller to stabilize the system \eqref{E1}.

\emph{\textbf{Experiment 3}}: Exerting one group of persistently exciting input signals on the  nominal system \eqref{E1}, and collecting input sequences and corresponding state sequences as
\begin{subequations}\label{E29}
\begin{align}
{U_ - } &= \left[ {\begin{array}{*{20}{c}}
{{u_0}}&{{u_1}}& \cdots &{{u_{T - 1}}}
\end{array}} \right] \tag{33a},\\
{X_ - } &= \left[ {\begin{array}{*{20}{c}}
{{x_0}}&{{x_1}}& \cdots &{{x_{T - 1}}}
\end{array}} \right] \tag{33b},\\
{X_ + } &= \left[ {\begin{array}{*{20}{c}}
{{x_1}}&{{x_2}}& \cdots &{{x_T}}
\end{array}} \right]. \tag{33c}
\end{align}
\end{subequations}

Certainly, a descriptor system can be persistently excited by input signals, but the full-row rank condition \eqref{E27} for normal systems is hard to be satisfied on descriptor systems' datasets $({U_ - },{X_ - })$. That is because the matrix $E$ is singular, so there are algebraic constraints between state signals, which means that some rows in ${X_ - }$ can be spanned by other remaining rows in ${X_ - }$ and ${U_ - }$. Thus, the condition \eqref{E27} is hard to be implemented for descriptor systems, which means that Lemmas \ref{Lem7} and \ref{Lem8} can not be directly applied.

However, inspired by the equivalent stabilizability of the nominal system \eqref{E1} and its slow subsystem \eqref{E2a}, if we can find a data-based decomposing method which transfers the nominal datasets $({U_ - },{X_ - })$ into its slow subsystem's datasets $({U_ - },{X_ - ^s})$, then a feedback controller $u_k = K_{s}x_k^s$ for the slow subsystem \eqref{E2a} can be easily obtained based on Lemmas \ref{Lem7} and \ref{Lem8}. Furthermore, after a reverse transformation to $K_s$ in \eqref{E26a}, a state feedback controller $u_k = Kx_k$ which stabilizes the nominal descriptor system \eqref{E1} can also be obtained. Luckily, such data-based decomposing method can be found in Experiment 1, specially in the equation \eqref{E13}.

\begin{lemma}  \label{Lem9}
For the singular matrix ${D_E} = M{N^{ - 1}} = {({s _0}E - A)^{ - 1}}E$ collected from data in Experiment 1, there exists a nonsingular matrix $\hat T\in \mathbb{R}^{n\times n}$, such that
\begin{equation} \label{E30}
{\hat T^{ - 1}}{D_E}\hat T = \texttt{diag}({{\hat E}_1},{{\hat E}_2})
\end{equation}
where ${{\hat E}_1}\in \mathbb{R}^{n_1\times n_1}$ is nonsingular and ${{\hat E}_2}\in \mathbb{R}^{n_2\times n_2}$ is nilpotent. Furthermore, by choosing nonsingular matrices
\begin{subequations}\label{E31}
  \begin{equation} \label{E31a}
    Q =\texttt{diag}(\hat E_1^{ - 1},{(s_0 {{\hat E}_2} - I)^{ - 1}}){\hat T^{ - 1}}{({s _0} E - A)^{ - 1}},
  \end{equation}
  \begin{equation} \label{E31b}
    P = \hat T,
  \end{equation}
\end{subequations}
the descriptor system \eqref{E1} can be transferred into its slow-fast subsystems' form \eqref{E2}.
\end{lemma}

\emph{Proof:} Based on the core-nilpotent decomposition of singular matrices, a nonsingular matrix $\hat T$ can always be found. Then choosing $Q$ and $P$ as \eqref{E31}, we can easily obtain
$QEP = \texttt{diag} ({I_{{n_1}}},{({s _0}{{\hat E}_2} - I)^{ - 1}}{{\hat E}_2})$ and $QAP = \texttt{diag}({s _0}{I_{{n_1}}} - \hat E_1^{ - 1},{I_{{n_2}}})$. It is clear that ${({s _0}{{\hat E}_2} - I)^{ - 1}}{{\hat E}_2}$ is nilpotent and $({s _0}{I_{{n_1}}} - \hat E_1^{ - 1})$ is nonsingular, so the proof is completed. \hfill $\square$

\begin{remark} \label{Rmk5}
Lemma \ref{Lem9} presents a data-based method to transfer a descriptor system into its slow-fast subsystems' form. Due to the unknown matrices $E$ and $A$, the matrix $Q$ is unavailable, but luckily the matrix $P$ can be obtained, and that is enough for us to access the state of slow subsystem from the nominal system. Although the matrix $P$ is not unique, as long as we find one through the core-nilpotent decomposition method in \eqref{E30} and \eqref{E31}, it can always transfer the nominal state into the slow-fast subsystems' state. To illustrate, supposing we obtain another $P^{'}$, then based on Lemma \ref{Lem5}, we have
\begin{equation} \label{E32}
x = P\begin{bmatrix}{x_k^s}\\
{x_k^f} \\\end{bmatrix} = {P^{'}}\begin{bmatrix}{T_1} & 0 \\0 & {T_2} \\\end{bmatrix}\left[ {\begin{array}{*{20}{l}}
{x_k^s}\\
{x_k^f}
\end{array}} \right] = {P^{'}}\begin{bmatrix}{{T_1}x_k^s}\\
{{T_2}x_k^f} \\\end{bmatrix}
\end{equation}
where $T_1\in \mathbb{R}^{n_1\times n_1}$ and $T_2\in \mathbb{R}^{n_2\times n_2}$ are nonsingular matrices, which means the nominal state $x$ can always be splitted into the slow-fast subsystem's state.
\end{remark}

To obtain a state feedback controller, a controllable normal system is required in Lemmas \ref{Lem7} and \ref{Lem8}, and such requirement is also needed for the slow subsystem \eqref{E2a}. For descriptor systems, slow subsystems are controllable, if and only if nominal systems are R-controllable \cite{dai1989singular}, which can be easily testified by Theorem \ref{Th4r}.
\begin{theorem} \label{Th5}
For a nominal R-controllable descriptor system \eqref{E1}, if the condition \eqref{E27} holds, where the datasets $({X_ - ^s},{X_ + ^s})$ are obtained from
\begin{subequations}\label{E33}
\begin{align}
{P^{ - 1}}{X_ - } &= \left[ {\begin{array}{*{20}{l}}
{X_ - ^s}\in \mathbb{R}^{n_1\times T} \\
\hline
{X_ - ^f}\in \mathbb{R}^{n_2\times T}
\end{array}} \right],  \tag{37a}\\
{P^{ - 1}}{X_ + } &= \left[ {\begin{array}{*{20}{l}}
{X_ + ^s}\in \mathbb{R}^{n_1\times T}\\
\hline
{X_ + ^f}\in \mathbb{R}^{n_2\times T}
\end{array}} \right], \tag{37b}
\end{align}
\end{subequations}
and $n_1$, $n_2$ and $P$ are obtained from Experiment 1 and Lemma \ref{Lem9}, then solving LMI conditions in \eqref{E28}, the feedback gain $K_s$ for the slow subsystem \eqref{E2a} can be obtained. Furthermore, a state feedback controller $u_k = K x_k$ which stabilizes the nominal descriptor system \eqref{E1} is given by $K = \left[ {\begin{array}{*{20}{c}}
{{K_s}}&0
\end{array}} \right]{P^{ - 1}}$.
\end{theorem}

\emph{Proof:} Based on the equivalent stabilizability of the nominal system \eqref{E1} and its slow subsystem \eqref{E2a} in Lemma \ref{Lem6}, plus Lemmas \ref{Lem7}, \ref{Lem8} and \ref{Lem9}, it is straightforward to obtain the results in Theorem \ref{Th5}. \hfill $\square$

\section{Simulation}\label{sec5}

\begin{figure}
\centering
\includegraphics[scale = 0.5]{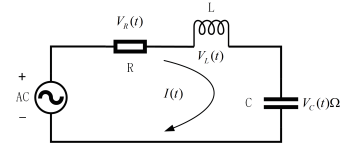}
\caption{A single-loop circuit network}
\label{F1}
\end{figure}

Consider the circuit network system \cite{dai1989singular} in figure \ref{F1}, by choosing the state $x = \begin{bmatrix}
I & V_L & V_C & V_R \\\end{bmatrix}^T$ and the control input $ u = V_S$, we can obtain a nominal descriptor system \eqref{E1}, where $E = \begin{bmatrix}L&0&0&0\\
0&0&1&0\\
0&0&0&0\\
0&0&0&0 \\\end{bmatrix}$, $A =\begin{bmatrix}
0&1&0&0\\
\frac{1}{C}&0&0&0\\
-R&0&0&1\\
0&1&1&1 \\\end{bmatrix}$ and $B = \begin{bmatrix}0 &0 &0&-1 \\\end{bmatrix}^T$. Here, we take $R, L$ and $C$ as $R = 1\Omega$, $L = 1H$ and $C=1F$ to generate data. Inputs in Experiments 1 and 3 and the initial system state $x_0$ satisfying Assumption \ref{Ass2} are given by MATLAB function \texttt{rand()}. Note that the nominal system is not in the slow-fast subsystems' form, so other existing methods are not applicable, and we will show the effectiveness of our methods. For all experiments, we choose $s_0 = 0.5$.

(\uppercase\expandafter{\romannumeral1}) System type identification: (1) No noises: Choosing the step length $l = 4$, after collecting data from Experiments 1, we have $\texttt{r}(M) = 2$. Thus, we conclude that $\texttt{r}(E) = 2$ and the system is a descriptor system. (2) System noises: Adding random noisy vectors within $\begin{bmatrix}-0.01 & 0.01 \\\end{bmatrix}$, and collecting data from Experiment 1. If we continue to compute the rank of $\hat{M}$, it comes to $\texttt{r}(\hat{M}) = 3$, which leads to a misjudgment. The Table \ref{tab1} shows the singular values of $\hat{M}$ with different $l$. It can be observed that comparing with $\sigma_1(\hat{M})$ and $\sigma_2(\hat{M})$, $\sigma_3(\hat{M})$ and $\sigma_4(\hat{M})$ are small scalars which can be attributed to noises, so we conclude that $\texttt{r}(E) = 2$. Besides, with a larger $l$, the difference between true non-zero singular values and the perturbations tends to be more evident.

\begin{table}[htbp]
\centering
\caption{Singular values of $\hat{M}$ with different $l$}
\begin{tabular}{lllll}
    \toprule
& $l=4$ & $l=100$ & $l=1000$ & $l=10000$ \\
    \midrule
  $\sigma_1(\hat{M})$ & 5.4166 & 90.4186 & 882.5501 & 8817.2461 \\
  $\sigma_2(\hat{M})$ & 0.6992 & 3.1251 & 11.6315 & 30.3656 \\
  $\sigma_3(\hat{M})$ & 0.0028 & 0.0029 & 0.0032 & 0.0046 \\
  $\sigma_4(\hat{M})$ & 0 & 0 & 0 & 0 \\
   \bottomrule
  \end{tabular}
 \label{tab1}
\end{table}

(\uppercase\expandafter{\romannumeral2}) Controllability test: Choosing the step length $l = 4$, after collecting data from Experiments 1 and 2, we have
\begin{equation}
\begin{split}
&\texttt{r} ( {\left[ {\begin{array}{*{20}{c}}
{{D_B}}&{{D_E}{D_B}}& {{{( {{D_E}})}^{2}}{D_B}} &{{{( {{D_E}})}^{3}}{D_B}}
\end{array}} \right]} ) = 3 \neq 4, \\
&\texttt{r}(\left[ {\begin{array}{*{20}{c}}
{D_E}&0\\
{D_A}&{D_E}
\end{array}} \right]) = 6, \texttt{r}(\left[ {\begin{array}{*{20}{l}}
{D_E}&0&0\\
{D_A}&{D_E}&{D_B}
\end{array}} \right]) = 6,\\
&\texttt{r}(M_D) = 16.
\end{split}
\nonumber
\end{equation}
Thus, we conclude that the system is a causal system, and it is causally controllable and R-controllable, but not completely controllable, which are consistent with MBC conclusions.

\begin{figure}
\centering
\includegraphics[scale = 0.5]{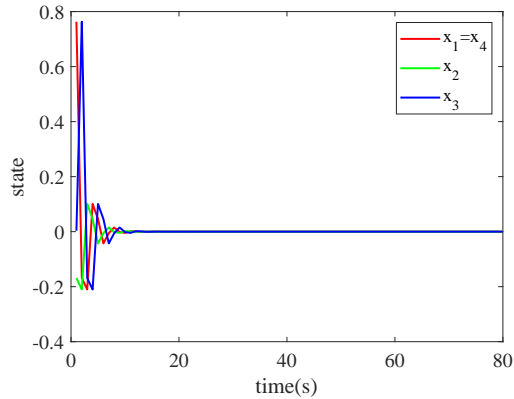}
\caption{State sequences of the closed-loop system}
\label{F2}
\end{figure}

(\uppercase\expandafter{\romannumeral3}) Stabilization: Choosing sampling time $t = 0.1s$ and length $T = 8s$, then we obtain datasets $U_{-}$, $X_{-}$ and $X_{+}$ from Experiment 3. After a core-nilpotent decomposition of singular matrices $D_E$ in Experiment 1, we obtain the decomposing matrix
$\hat T= P = \left[ {\begin{array}{*{20}{c}}
{-0.2162}&{0.5944}&0&0\\
{ - 0.5564}&{ - 0.5389}&-0.8455&0.5340\\
0.7726&-0.0554&0&0\\
{-0.2162}&{0.5944}&0.5340&0.8455
\end{array}} \right]$. Furthermore, with a transformation \eqref{E33} to collected datasets $(X_ -,X_ +)$, the datasets $(X_ - ^s,X_ + ^s)$ are obtained. Then solving the LMI condition in Lemma \ref{Lem8} by CVX \cite{grant2014CVX}, the controller for the slow subsystem \eqref{E2a} is given as
${K_s} = \left[ {\begin{array}{*{20}{l}}
{0.6982}&{0.2489} \end{array}} \right]$. Finally, the state feedback controller for the nominal system \eqref{E1} is obtained as
$K = \left[ {\begin{array}{*{20}{c}}
{{K_s}}&0
\end{array}} \right]{P^{ - 1}} = \left[ {\begin{array}{*{20}{l}}
{0.5165}&0&1.0482&0\end{array}} \right]$. It can be checked that all eigenvalues of matrix pencil $(E,A + BK)$ are in the unite circle. Besides, figure \ref{F2} shows the state sequences of the closed-loop system, which verifies that the system is stabilized.

\section{Conclusions}\label{sec6}

In this paper, three data experiments are designed to testify a parameter-unknown descriptor system's type, analyze its controllability as well as stabilize it with a state feedback controller. On one hand, our methods can help researchers interested in DDC control for normal systems check whether their common assumption that the system is normal holds or not, so that their methods will not be misused. On the other hand, if the system is a descriptor one, its properties can be well revealed by our methods. More importantly, a data-based decomposing method is proposed to transfer the descriptor system to its slow-fast subsystems' form. In this way, due to the equivalent stabilizability between the  nominal system and its slow subsystem, a state feedback controller can be designed to stabilize such a system. In the future, we will investigate the numerical stability of our methods in the presence of uncertainties and disturbances. Besides, other control requirements such as LQG and ${H_\infty }$ will also be considered.

\section*{Acknowledgment}\label{sec7}

This work was supported by the National Natural Science Foundation of China (No.62003186 and No.62103225), the Natural Science Foundation of Guangdong, China (No.2020A-1515010334 and No. 2021A1515012628), and the Shenzhen Science and Technology Program, China (No.JCYJ2021-0324132606015).
\bibliographystyle{IEEEtran}
\bibliography{IEEE-J}

\end{document}